# Waveguiding-assisted random lasing in epitaxial ZnO thin film

P.-H. Dupont [a)], C. Couteau [a)*], D. J. Rogers [b)], F. Hosseini Téhérani [b)], and G. Lérondel [a)]

a)  Laboratoire de Nanotechnologie et d'Instrumentation Optique, Institut Charles Delaunay, Université de Technologie de Troyes, 12 rue Marie Curie, 10000 Troyes, France.

b)  Nanovation SARL, 103bis rue de Versailles, 91400 Orsay, France

Zinc Oxide thin films were grown on c-sapphire substrates using pulsed laser deposition. Pump power dependence of surface emission spectra, acquired using a quadrupled 266 nm laser, revealed room temperature stimulated emission (threshold of 900 kW/cm$^2$). Time dependent spectral analysis plus gain measurements of single-shot, side-emission, spectra pumped with a nitrogen laser revealed random lasing indicative of the presence of self-forming laser cavities. It is suggested that random lasing in an epitaxial system rather than a 3-dimensional configuration of disordered scattering elements, was due to waveguiding in the film. Waveguiding causes light to be amplified within randomly-formed closed-loops acting as lasing cavities.

* Corresponding author: christophe.couteau@utt.fr

Properties of Zinc Oxide (ZnO) are the subject of intense research [1] and many optoelectronic applications are emerging. ZnO is interesting as it has intense excitonic ultraviolet (UV) stimulated emission, due to high exciton binding energy (60 meV) and a direct wide band-gap of ~3.37 eV [1].

We studied random laser action, observed in many different forms of ZnO [2]. A random laser has no deliberately-formed cavity and light amplification occurs due to scattering at dielectric reflectors. With enough scatterers and optical gain, light can reflect back to its initial position, forming closed-loops acting as self-forming laser cavities. Random lasing was seen for GaAs, ZnO and $TiO_2$ and reported in powders [3], dyes embedded in films [4] and in highly textured [5] and non-oriented polycrystalline films [6]. Electrical pumping was also achieved [7]. In contrast to these results, we report random lasing in an epitaxial thin film.

In this study, ZnO thin films were grown on c-sapphire (c-$Al_2O_3$) substrates using pulsed laser deposition (PLD). High resolution x-ray diffraction (HRXRD) studies were conducted in a Panalytical X-Pert MRD PRO system using a Cu $K\alpha$ source. The incident beam optics comprised a four-bounce Ge (220) monochromator and a multilayer mirror. In the diffracted beam path, the system was equipped with a three-bounce Ge (220) monochromator. Optical emission was investigated using two geometries: in the first, a frequency-quadrupled neodymium-doped yttrium aluminium garnet (Nd:YAG) laser (266 nm, 5 ns pulse duration, 10 Hz frequency and 4 mm spot diameter) was employed as a pump, with the beam having an incidence angle of 45° to the sample surface. The emission was collected normal to the sample surface using a spectrometer (50 cm focal length) coupled to a Peltier-cooled CCD camera. The second "edge-emission" geometry used a nitrogen pump laser emitting at 337 nm (5 ns pulse duration, 10 Hz frequency) with the incident beam normal to the sample surface and the collection at the edge of the sample (at 90° from the excitation beam: see inset in Fig. 4).

In the edge-emission configuration, a cylindrical lens with a 35 mm focal length was used to create a stripe (width of 30 µm), with a power density of up to 6.6 MW/cm². The CCD camera acquisition time was fixed at 100 ms in order to obtain "single-shot spectra" (corresponding to the 10 Hz frequency of the excitation laser).

Gain measurements were taken using the Variable Stripe-Length (VSL) method. In this approach, the laser stripe length is defined using a movable beam-block. Side-emission intensity is measured as a function of the stripe length. This allows for the excitation volume to be increased whilst maintaining a constant pump power density.

X-ray diffraction (XRD) measurements revealed epitaxial growth [8,9] with a FWHM of 0.006° for the (0002) peak high resolution omega scan and an (open-detector) omega rocking curve value of about 0.05°, characteristic of a highly oriented material with very low crystallographic dispersion. The 2θ-ω scan of the (0002) peak gave a c lattice parameter of 5.215 Å. This is slightly higher than the equilibrium value suggesting the film was under compressive stress in the basal plane. The 2θ-ω scan also showed strong Pendellösung oscillations, characteristic of a high parallelicity of the crystal planes combined with a smooth sample surface over a relatively large area compared with the laser spot size (several mm$^2$). Fringe spacing gave a film thickness of 140 nm, as compared to a value of 159 nm calculated from angularly-resolved photoluminescence (PL) and spectroscopic ellipsometry measurements (not shown).

For surface emission PL spectra, Fig. 1(a) presents the evolution of the luminescence as a function of the pump power for an accumulation of 50 single-shot spectra at various power densities. The inset shows the integrated emission intensity as a function of pump power. A stimulated emission threshold of 900 kW/cm$^2$ was identified and is a typical value for an epitaxial ZnO thin film [10]. For a pump power of 400 kW/cm$^2$, spontaneous emission is observed at 378 nm due to near band edge (NBE) recombination. Above the stimulated emission threshold, however, the main peak is over ~394 nm, most likely due to electron-hole plasma recombination [11]. With increasing pump power, peak broadening and red shifting are observed. This is attributable to band-gap renormalization [12]. Fig. 1(b) presents a series of single-shot spectra, for which the power density was just above the stimulated emission threshold (1.1 MW/cm$^2$). The PL intensity was observed to fluctuate significantly between spectra on a random basis over time.

Figure 2(a) shows groups of five single-shot side-emission spectra acquired at various pump powers. For a pump power of 3.7 MW/cm$^2$ (above threshold), individual spectra show multiple emission peaks, with each having significantly lower linewidth than observed for the surface emission spectra (at similar pump powers). This is consistent with the presence of laser cavities in the sample, which select precisely defined wavelengths. Moreover, the various emission peaks were observed to toggle on and off randomly over time (as was also observed for the surface emission), which is coherent with varying lasing modes being excited by each pulse. Thus it was concluded that random lasing occurs in the film, with different self-forming laser cavities being selected at each pump pulse. Figure 2(b) shows 50 spectra juxtaposed for different pump powers (ranging from 2 MW/cm$^2$ to 6.7 MW/cm$^2$), as a contour plot. We see that although the peaks in the single-shot spectra occur randomly, they appear at well-defined positions, which recur over time. Figure 3 shows the result of accumulating 50 side-emission spectra at various pump powers. The inset shows the dependence of the spectrally integrated emission intensity as a function of the pump power. A stimulated emission threshold of 2.9 MW/cm² was deduced. We note that spontaneous emission, observed sub-threshold, appeared at a higher wavelength (i.e. 390 nm) than for the normal collection case. This may be due to waveguiding in the film, which would favour TE over TM polarisation mode at lower wavelengths. At higher pump powers, the output begins to saturate. Moreover, the stimulated emission thresholds are slightly different for the surface and edge emission. This may be the result of using a stripe laser (length of 4 mm, depth of 200 μm, i.e. more than the film thickness) combined with a cylindrical lens for the side configuration.

Using VSL measurements (Fig. 4), lasing threshold was found to depend not only on the laser pump power but also on the excitation volume, as was concluded in studies on random lasers elsewhere [13]. As the excitation volume increases beyond a threshold of 482 μm$^3$ (a stripe of length 90 μm, depth 159 nm), laser action takes place. However, peak separation was not observed to vary as a function of stripe length which distinguishes these results from those typically observed (e.g. Tang *et al.* [14]), suggesting our findings should not be interpreted in the same way. The peaks had widths between 0.3 and 2 nm for single-shot spectra and a gain, g, of 160 cm$^{-1}$ (c.f. Fig. 4), coherent with the literature for comparable samples [15].

Light emitted was polarized, with all peaks exhibiting a maximum corresponding to the TE mode (inset, Figure 4). This indicates the ZnO film may act as a waveguide, as observed before [14], and is also consistent with the red-shift observed for the spontaneous emission in side-collection.

Having an epitaxial thin film poses questions on the origin of the random lasing. Figure 5 shows a simulation for a Fabry-Perot cavity spectrum fitted to these results (assuming Gaussian gain, including wavelength dispersion caused by the change of refractive index in ZnO). Fitting parameters included maximum gain amplitude, $g_{max}$, of 471 cm$^{-1}$, 1% reflectivity and a cavity length of 12.03 µm. We stress that $g_{max}$ is a non-adjustable parameter as a gain of 160 cm$^{-1}$ was measured by VSL (see Fig. 4) (i.e. the Gaussian fit for the gain was made for a FWHM of 160 cm$^{-1}$, resulting in $g_{max}$ of 471 cm$^{-1}$). This simulation fits the experimental results fairly well meaning we have scattering due to reflection at defects in the film. We point out the deduced "cavity length" (12.03 µm) does not necessarily imply there are actual cavities of this dimension in the sample, but rather that the typical path of the photon before arriving back at its original position is of the order of 12 µm. While such a one-dimensional Fabry-Perot model can be considered as over-simplified, the probability for a photon to return to its emitted point can be deduced from the reflectivity to be 0.01. As far as the scattering process is concerned, it has been reported that space charge effect at grain boundaries increases the refractive index of the medium locally, therefore acting as a mirror for the light [16]. This hypothesis needs confirmation using high resolution transmission electron microscopy observations.

In conclusion, room temperature excitation dependence of optical gain spectra was investigated for thin films of ZnO epitaxially grown on c-sapphire substrates by PLD. Time dependent analysis of single-shot side-emission spectra using a VLS approach revealed room temperature random lasing from self-forming cavities, with a threshold of 2.9 MW/cm². It was proposed that waveguiding in the film enhanced the random lasing. The origin of scattering remains uncertain. Such an experimental approach could be used as a tool to characterise thin film quality via random lasing observations.

We acknowledge financial support by the French Agence National pour la Recherche (ANR) project UltraFlu ANR-08-BLAN-0296 and we thank Prof. Hui Cao and Prof. Heinz Kalt for helpful discussions.

**Figure captions**

FIGURE 1: a- Single-shot surface emission PL spectra. Inset: experimental configuration used. b- Accumulated spectra for 50 surface emission pulses at different pump powers. Inset: Integrated emission intensities at different pump powers. The threshold is at 0.9 MW / cm².

FIGURE 2: a- Several single-shot side-emission spectra at various pump powers.
b- 3D Contour plot for single-shot side-emission spectra at different pump powers. The y-axis represents the wavelength while the x-axis represents the pump power from the right (2 MW/cm$^2$) to the left (6.7 MW/cm$^2$). Each section of pump power contains 50 spectra side by side. The z-axis represents the intensity, which increases from low intensity (black) to high intensity (white), whilst going through intermediate intensity (grey). Note the recurrence of certain peak positions at higher powers.

FIGURE 3: Averaged spectra for 50 side-emission pulses at different pump powers. Inset: Integrated emission intensities as a function of pump power. The stimulated emission threshold is 2.9 MW / cm².

FIGURE 4: Side emission VSL gain measurements (bottom inset). Top inset: Emission intensity versus angle of polarization. The maximum corresponds to the TE mode.

FIGURE 5: Simulated Fabry-Perot oscillator (cavity length: 12.03µm, Reflectivity: 1%) with Gaussian gain (max: 471 cm$^{-1}$) fitted against averaged PL spectra at 6.6 MW / cm² pump intensity.

Figure 1

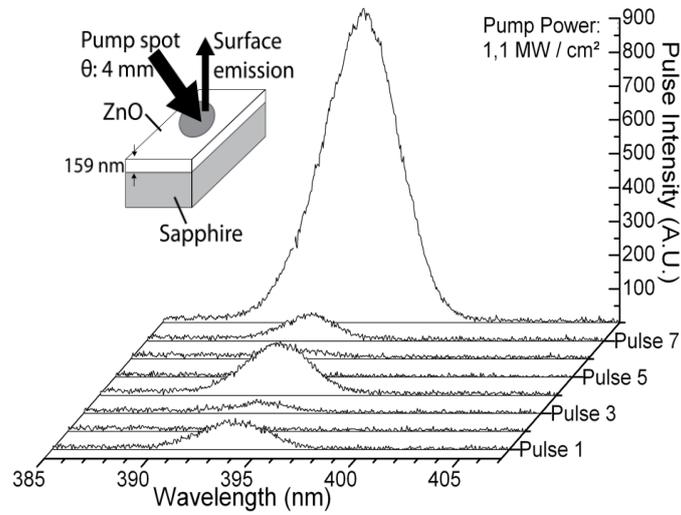

a)

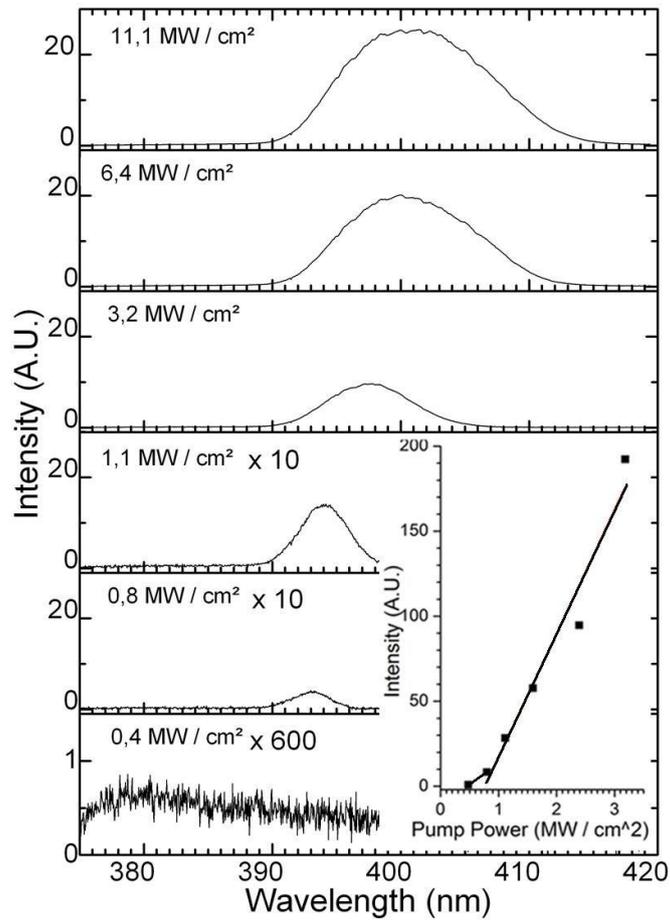

b)

Figure 2

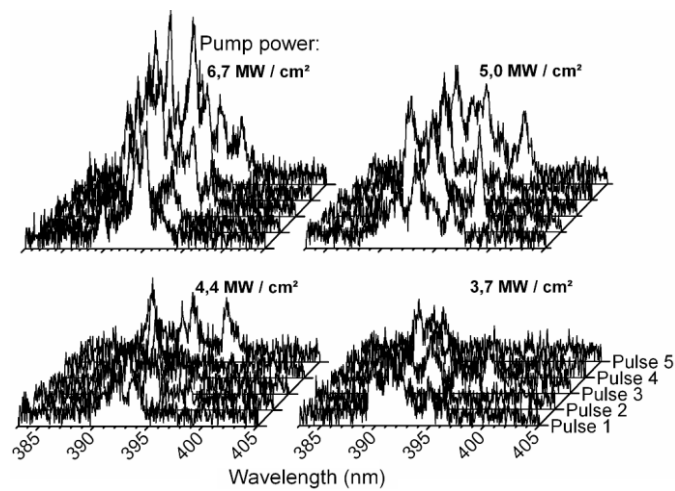

a)

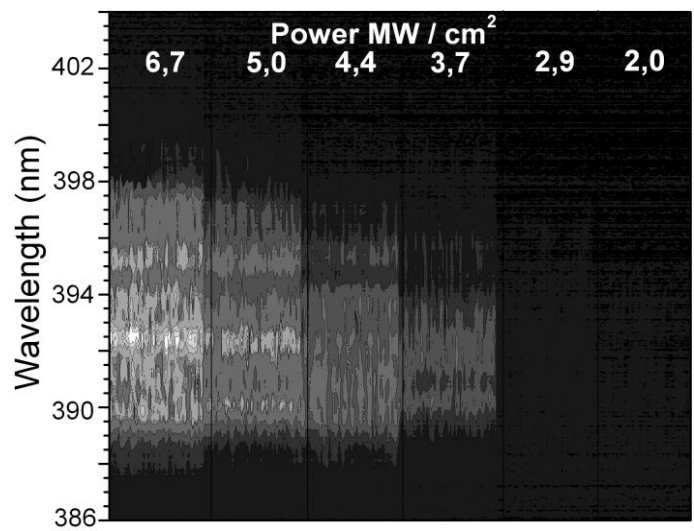

Figure 3

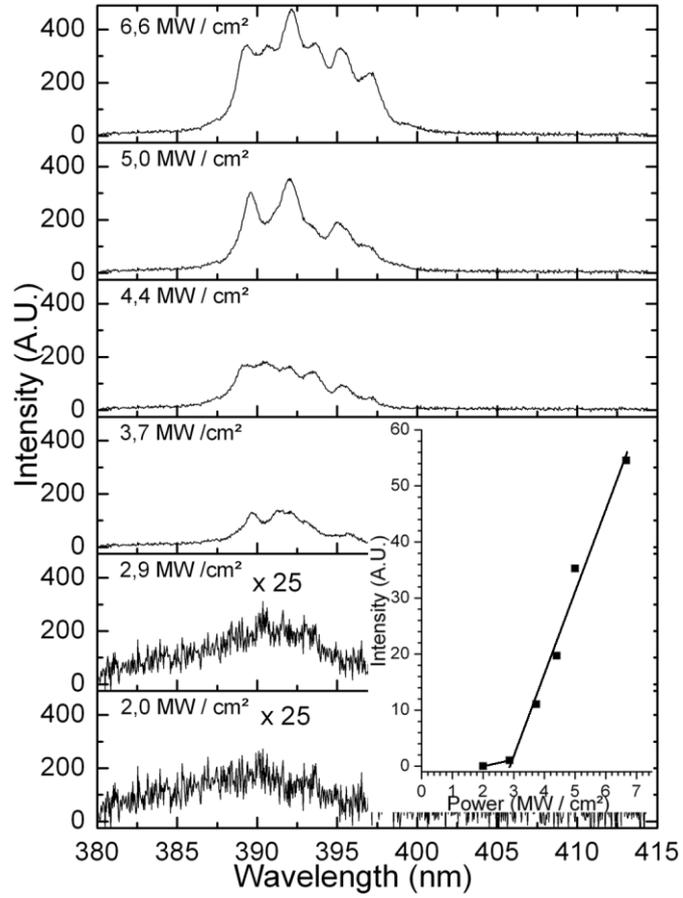

Figure 4

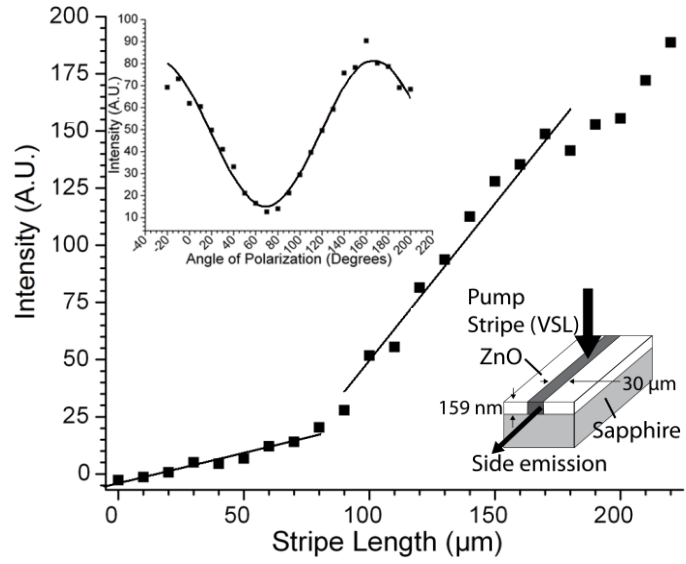

Figure 5

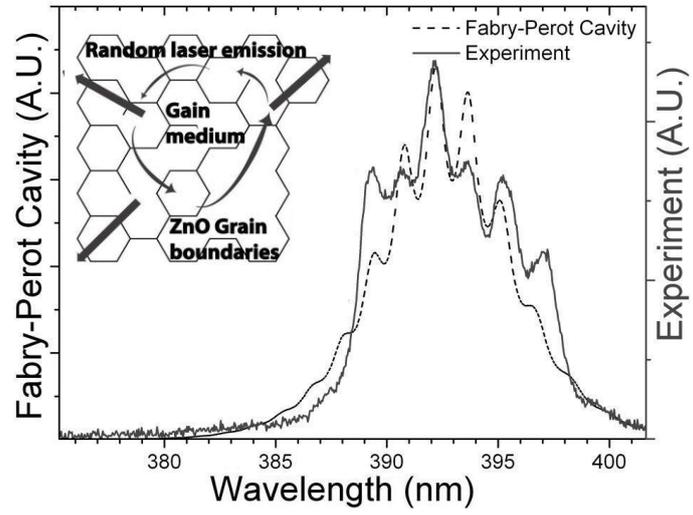